\documentclass[preprint,12pt]{aastex}

\shortauthors{Winn et al.~2008}
\shorttitle{Transits of WASP-4b}

\begin{document}

%
\def\ltsima{$\; \buildrel < \over \sim \;$}
\def\lsim{\lower.5ex\hbox{\ltsima}}
\def\gtsima{$\; \buildrel > \over \sim \;$}
\def\gsim{\lower.5ex\hbox{\gtsima}}
%

\bibliographystyle{apj}

\title{
  The Transit Light Curve Project.\\
  XI.~Submillimagnitude Photometry of Two Transits \\
      of the Bloated Planet WASP-4b$^1$
}

\author{
Joshua N.\ Winn\altaffilmark{2},
Matthew J.\ Holman\altaffilmark{3},
Joshua A.\ Carter\altaffilmark{2},\\
Guillermo Torres\altaffilmark{3},
David J.\ Osip\altaffilmark{4},
Thomas Beatty\altaffilmark{2}
}

\altaffiltext{1}{Based on observations with the 6.5m Magellan
  Telescopes located at Las Campanas Observatory, Chile.}

\altaffiltext{2}{Department of Physics, and Kavli Institute for
  Astrophysics and Space Research, Massachusetts Institute of
  Technology, Cambridge, MA 02139, USA}

\altaffiltext{3}{Harvard-Smithsonian Center for Astrophysics, 60
  Garden Street, Cambridge, MA 02138, USA}

\altaffiltext{4}{Las Campanas Observatory, Carnegie Observatories,
  Casilla 601, La Serena, Chile}

\begin{abstract}

  We present photometry of two transits of the giant planet WASP-4b
  with a photometric precision of 400--800~parts per million and a
  time sampling of 25--40~s. The two midtransit times are determined
  to within 6~s. Together with previously published times, the data
  are consistent with a constant orbital period, giving no compelling
  evidence for period variations that would be produced by a satellite
  or additional planets. Analysis of the new photometry, in
  combination with stellar-evolutionary modeling, gives a planetary
  mass and radius of $1.237\pm 0.064$~$M_{\rm Jup}$ and $1.365\pm
  0.021$~$R_{\rm Jup}$. The planet is 15\% larger than expected based
  on previously published models of solar-composition giant
  planets. With data of the quality presented here, the detection of
  transits of a ``super-Earth'' of radius $1.75$~$R_\oplus$ would have
  been possible.

\end{abstract}


\section{Introduction}

Wilson et al.~(2008) recently reported the discovery of WASP-4b, a
giant planet that orbits and transits a G7V star with a period of
1.34~days. This discovery is notable because the planet has an
unusually large radius and short orbital period, and the star is one
of the brightest transit hosts that is known in the Southern sky
($V=12.5$). The large radius seems to place the planet among the
``bloated'' planets for which there is no clear explanation (see,
e.g., Burrows et al.~2007, Guillot 2008). The short period raises the
possibility of observing tidal decay (Rasio et al.~1996, Sasselov
2003) and makes WASP-4b an attractive target for observations of
occultations (secondary eclipses) that would lead to detections of the
reflected light and thermal emission from the planet's atmosphere.

The host star's brightness and southern declination are important for
a practical reason: they allow the large-aperture telescopes of the
southern hemisphere to be used advantageously. In this paper, we
report on observations of a transit of WASP-4b with the Magellan/Baade
6.5m telescope, with the goal of deriving independent and refined
parameters for this interesting system. Previous papers in this
series, the Transit Light Curve project, have achieved this goal by
combining the information from many independent transit observations
with smaller telescopes (see, e.g., Holman et al.~2006, Winn et
al.~2007). In principle, with a larger telescope, it should be
possible to achieve this goal with fewer observations and also to
measure precise midtransit times, which can be used to search for
additional planets via the method of Holman \& Murray (2005) and Agol
et al.~(2005). Gillon et al.~(2008) recently presented photometry of
WASP-4 with one of the 8.2m Very Large Telescopes, with the same
motivation.

This paper is organized as follows. \S~2 describes the observations
and data reduction, \S~3 describes the photometric analysis, \S~4
describes the results of stellar-evolutionary modeling, and \S~5
discusses the newly-measured midtransit times and a refined
ephemeris. Finally, \S~6 discusses the refined measurement of the
planetary radius, and considers how small a planet we could have
detected, given data of the quality presented here.

\section{Observations and Data Reduction}

We observed the transits of UT~2008~Aug~19 and 2008~Oct~09 with the
Baade~6.5m telescope, one of the two Magellan telescopes at Las
Campanas Observatory in Chile. We used the Raymond and Beverly Sackler
Magellan Instant Camera (MagIC) and its SITe $2048\times 2048$~pixel
CCD detector, with a scale of $0\farcs069$~pixel$^{-1}$. Ordinarily
this detector uses four amplifiers, each of which reads a quadrant of
the array, giving a total readout time of 23~s. We used a $2048\times
256$~pixel subarray and a single amplifier, giving a readout time of
10~s. We rotated the field of view to align WASP-4 and a nearby
comparison star along the long axis of the subarray (parallel to the
read register). The comparison star is $36\arcsec$ east and
$71\arcsec$ south of WASP-4. At the start of each night, we verified
that the time stamps recorded by MagIC were in agreement with
GPS-based times to within one second. On each night we obtained
repeated $z$-band exposures of WASP-4 and the comparison star for
about 5~hr bracketing the predicted transit time. Autoguiding kept the
image registration constant to within 10~pixels over the course of the
night.

During the 2008~Aug~19 observations, WASP-4 rose from an airmass of
1.09 to 1.03, and then set to an airmass of 1.23. At first, we used an
exposure time of 30~s. Shortly after midtransit the seeing improved
abruptly, from a full-width at half-maximum (FWHM) of 11 pixels to 7
pixels. As a result of the higher rate of detected photons per pixel,
some images were spoiled due to nonlinearity and saturation. For the
rest of the night we used an exposure time of $t_{\rm exp} = 15$~s.

During the 2008~Oct~09 observations, WASP-4 rose from an airmass of
1.10 to 1.03, and then set to an airmass of 1.21. Having learned our
lesson in August, we deliberately defocused the telescope so that the
image width was dominated by the effects of the telescope aberration
rather than natural seeing. This was done by moving the secondary
mirror by a constant displacement relative to the in-focus position
determined by the guider probe. The exposure time was $t_{\rm exp} =
30$~s. The stellar images were ``donuts'' with a diameter of
approximately 25~pixels.

We used standard IRAF procedures for overscan correction, trimming,
bias subtraction, and flat-field division. The bias frame was
calculated from the median of 60~zero-second exposures, and the flat
field for each night was calculated from the median of 60~$z$-band
exposures of a dome flat screen. We performed aperture photometry of
WASP-4 and the comparison star and divided the flux of WASP-4 by the
flux of the comparison star. We experimented with different aperture
sizes and sky regions, aiming to minimize the variations in the
out-of-transit (OOT) portion of the differential light curve. Best
results were obtained with an aperture radius of 38~pixels for the
2008~Aug~19 observations and 35~pixels for the 2008~Oct~09
observations.

\begin{figure}[p]
\epsscale{0.9}
\plotone{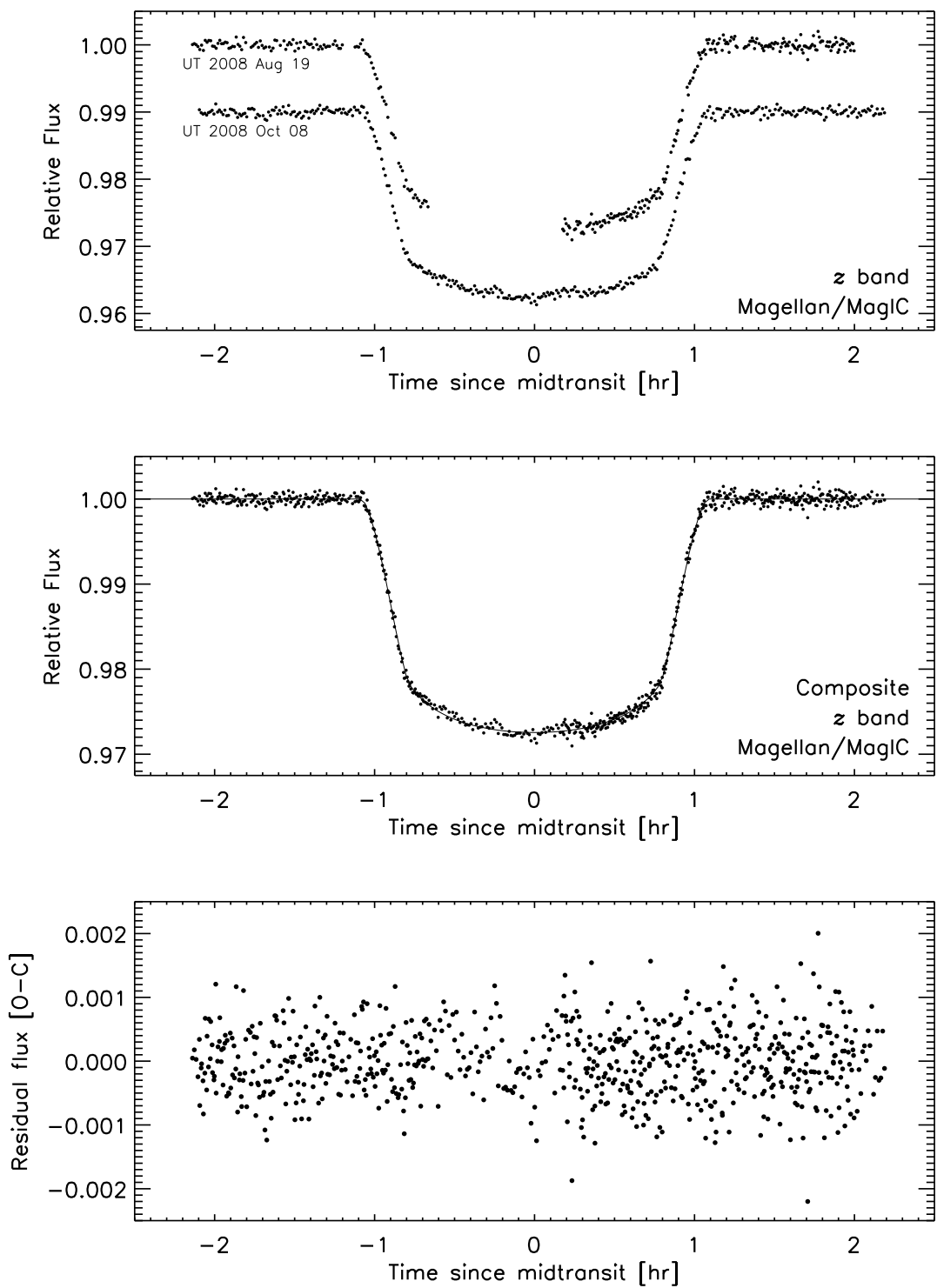}
\caption{
{\it Top.}---Relative $z$-band photometry of WASP-4 based on
  observations with the Magellan (Baade) 6.5m telescope.
{\it Middle.}---Composite light curve. The solid line shows the best-fitting
model.
{\it Bottom.}---Residuals between the data and the best-fitting model.
\label{fig:phot}}
\end{figure}

A few images were also obtained with Johnson-Cousins $VRI$ filters to
measure the difference in color between WASP-4 and the comparison
star. The instrumental magnitude differences (target minus WASP-4)
were $\Delta V = 0.434$, $\Delta R = 0.469$, $\Delta I = 0.478$, and
$\Delta z = 0.472$. Evidently the two stars are similar in color, with
the comparison star being slightly bluer [$\Delta(V-I)=-0.044$]. As
described in \S~3, the $z$-band time series was corrected for
differential extinction between the target and comparison star by
fitting a linear function of airmass to the magnitude difference. The
results were consistent with the expectation that the bluer comparison
star suffers from greater extinction per unit airmass. The
extinction-corrected data are given in Table~\ref{tbl:phot}, and
plotted in Fig.~\ref{fig:phot}, along with the best-fitting model.

Due to the abrupt seeing change on 2008~Aug~19 and the associated
change in exposure level, we consider the 15~s and 30~s exposures as
two separate time series (TS1 and TS2). Thus, together with the
2008~Oct~09 time series (TS3), there were 3 time series to be
analyzed. The TS1 pre-ingress data has a median time sampling of 41~s
and a standard deviation of 478~parts per million (ppm). The TS2
post-egress data has a median time sampling of 26~s and a standard
deviation of 691~ppm. These noise levels are about 13\% larger than
the calculated noise due to photon-counting statistics, read noise,
sky noise, and scintillation noise (using the approximate formulas of
Reiger~1963 and Young~1967). The ratio of the observed TS1 noise to
the observed TS2 noise is 0.69, which is nearly equal to $(t_{\rm
  exp,2}/t_{\rm exp,1})^{-1/2} = 0.71$. The near-equality is evidence
that the dominant noise source is a combination of photon noise and
scintillation noise, both of which vary as $t_{\rm exp}^{-1/2}$. In
the final version of TS3, which has a median time sampling of 41~s,
the pre-ingress data has an rms of 475~ppm and the post-egress data
has an rms of 488~ppm. These values are about 17\% above the
calculated noise level. In neither light curve did we detect
signficant correlation between the noise and the pixel coordinates,
FWHM, or shape parameters of the stellar images.

\begin{deluxetable}{ccc}

\tabletypesize{\scriptsize}
\tablecaption{Relative Photometry of WASP-4\label{tbl:phot}}
\tablewidth{0pt}

\tablehead{
\colhead{Barycentric Julian Date} &
\colhead{Relative flux} &
\colhead{Uncertainty}
}

\startdata
 $  2454697.708386$  &  $  1.00005$  &  $  0.00050$ \\
 $  2454697.708861$  &  $  1.00018$  &  $  0.00050$ \\
 $  2454697.709337$  &  $  1.00003$  &  $  0.00050$ \\
 $  2454697.709810$  &  $  0.99975$  &  $  0.00050$ \\
 $  2454697.710285$  &  $  0.99930$  &  $  0.00050$ \\
 $  2454697.710759$  &  $  0.99996$  &  $  0.00049$ \\
 $  2454697.711234$  &  $  0.99917$  &  $  0.00050$ \\
 $  2454697.711708$  &  $  1.00067$  &  $  0.00050$ \\
 $  2454697.712185$  &  $  1.00039$  &  $  0.00050$ \\
 $  2454697.712657$  &  $  1.00066$  &  $  0.00049$ \\
 $  2454697.713132$  &  $  1.00062$  &  $  0.00050$ \\
 $  2454697.713606$  &  $  0.99982$  &  $  0.00050$ \\
 $  2454697.714082$  &  $  0.99976$  &  $  0.00049$
\enddata

\tablecomments{The time stamps in Column 1 represent the Barycentric
  Julian Date at the time of midexposure. The uncertainty in Column 3
  is the quantity $\sigma_{f,i}$ discussed in \S~3. We intend for this
  Table to appear in entirety in the electronic version of the
  journal. An excerpt is shown here to illustrate its format. The data
  are also available from the authors upon request.}

\end{deluxetable}

\section{Photometric Analysis}

We fitted a transit light curve model to the data, based on the
analytic formulae of Mandel \& Agol (2002). The set of model
parameters included the planet-to-star radius ratio $(R_p/R_\star)$,
the ratio of the stellar radius to the orbital semimajor axis
$(R_\star/a)$, the orbital inclination ($i$), and the midtransit time
$(T_c$). We also fitted for three parameters ($\Delta m_0$, $k_z$ and
$k_t$) specifying a correction for systematic errors,
\begin{equation}
\label{eq:diff-ext}
\Delta m_{\rm cor} = \Delta m_{\rm obs} + \Delta m_0 + k_z z + k_t t,
\end{equation}
where $z$ is the airmass, $t$ is the time, $\Delta m_{\rm obs}$ is the
observed magnitude difference between the target and comparison star,
and $\Delta m_{\rm cor}$ is the corrected magnitude difference that is
compared to the idealized transit model. The $k_z$ term specifies the
differential airmass extinction correction that was mentioned in
\S~2. The $k_t$ term was not strictly necessary to fit the data (it
was found to be consistent with zero), but we included it in order to
derive conservative error estimates for the transit times, as the
parameter $k_t$ is covariant with the transit time. We allowed $\Delta
m_0$, $k_t$, and $k_z$ to be specific to each time series, with the
exception that TS1 and TS2 had a common value of $k_z$ because those
data were obtained on the same night.

The final two model parameters were the coefficients ($u_1$ and $u_2$)
of a quadratic limb-darkening law,
\begin{equation}
\frac{I_\mu}{I_1} = 1 - u_1(1-\mu) - u_2(1-\mu)^2, 
\end{equation}
where $\mu$ is the cosine of the angle between the line of sight and
the normal to the stellar photosphere, and $I_\mu$ is the specific
intensity. We allowed $u_1$ and $u_2$ to vary freely subject only to
the conditions $u_1+u_2<1$, $u_1+u_2>0$, and $u_1>0$. It proved
advantageous to perform the fit using the linear combinations
\begin{equation}
v_1 = u_1 + 2.33~u_2,~~v_2 = u_1 - 2.33u_2,
\end{equation}
because $v_1$ and $v_2$ have nearly uncorrelated errors (for further
discussion, see P\'al~2008).

We assumed the orbit to be circular because the RV data of Wilson et
al.~(2008) are consistent with a circular orbit, and because the
expected timescale for orbital circularization at present,
\begin{equation}
\tau_c = \frac{4}{63}~Q_p~\left(\frac{P}{2\pi}\right)
\left(\frac{M_p}{M_\star}\right)
\left(\frac{a}{R_p}\right)^5,
\end{equation}
(Rasio et al.~1996) is 0.3~Myr~$(Q_p/10^5)$ for WASP-4b, which is much
shorter than the estimated main-sequence age of the star. In this
expression, $Q_p$ is the the tidal dissipation parameter (see, e.g.,
Goldreich \& Soter 1966). This order-of-magnitude argument suggests
that assuming a circular orbit is reasonable, although the value of
$Q_p$ for irradiated giant planets is highly uncertain, and the
expected timescale is highly approximate because it ignores the
coupled evolution of the orbital distance and eccentricity (Jackson,
Greenberg, \& Barnes 2008).

The fitting statistic was
\begin{equation}
\chi^2_F = \sum_{i=1}^{N} \left[ \frac{f_i({\rm obs}) - f_i({\rm calc})}{\sigma_{f,i}} \right]^2,
\label{eq:chi2_flux}
\end{equation}
where $N$ is the number of flux measurements (photometric data
points), $f_i$(obs) is the $i$th measurement, $f_i$(calc) is the
calculated flux given a particular choice of model parameters, and
$\sigma_{f,i}$ is the uncertainty in the $i$th measured flux. We
determined appropriate values of $\sigma_{f,i}$ as follows. First, we
multiplied the calculated errors in each time series by a constant
chosen to give a minimum value of $\chi_F^2/N_{\rm dof} = 1$. The
constants were 1.20, 1.12, and 1.21 for TS1, TS2, and TS3
respectively. Next, we assessed the time-correlated noise (also called
``red noise'') by examining the autocorrelation function, the power
spectrum, and a plot of the Allan (1966) deviation of the
residuals. We also used the method described by Winn et al.~(2008), in
which the ratio $\beta$ is computed between the standard deviation of
time-averaged residuals, and the standard deviation one would expect
assuming white noise. For TS1 and TS2 we found no evidence for
significant correlations. For TS3 we found structure in the
autocorrelation function on a time scale of 15-20~min, the approximate
ingress or egress duration, giving $\beta=1.52$. One naturally
suspects that the correlated noise represents measurement error,
although it is also possible that the noise is astrophysical, arising
from starspots or other stellar inhomogeneities. In support of an
astrophysical origin, we find no evidence for correlated noise
($\beta=1$) when considering only the out-of-transit
data. Nevertheless we cannot draw a firm conclusion; instead we
attempt to account for the correlations by multiplying the error bars
of TS~3 by an additional factor of $\beta=1.52$. Thus, for TS1, TS2,
and TS3, the final values of $\sigma_{f,i}$ were equal to the
calculated error bars multiplied by 1.20, 1.12, and $1.21\times
1.52=1.84$, respectively. The error bars given in Table~1 are the
final values of $\sigma_{f,i}$ that were used in the fitting process.

To determine the allowed ranges of each parameter, we used a Markov
Chain Monte Carlo (MCMC) technique, with the Gibbs sampler and the
Metropolis-Hastings algorithm, to estimate the {\it a posteriori}\,
joint probability distribution of all the model parameters. This
algorithm creates a sequence of points (a ``chain'') in parameter
space by iterating a jump function, which in our case was the addition
of a Gaussian random deviate to a randomly-selected single
parameter. After this operation, if the new point has a lower
$\chi_F^2$ than the previous point, the ``jump'' is executed: the new
point is added to the chain. If not, then the jump is executed with
probability $\exp(-\Delta\chi_F^2/2)$. When the jump is not executed,
the current point is repeated in the chain. The sizes of the random
deviates are adjusted so that $\sim$40\% of jumps are executed. After
creating 10 chains of 500,000 links to check for mutual convergence,
and trimming the first 20\% of the links to eliminate artifacts of the
initial conditions, the density of the chains' points in parameter
space was taken to be the joint {\it a posteriori}\, probability
distribution of the parameter values. Probability distributions for
individual parameters are created by marginalizing over all other
parameters.

The results are given in Table~\ref{tbl:params}, which gives the
median of each distribution, along with the 68.3\% lower and upper
confidence limits (defined by the 15.85\% and 84.15\% levels of the
cumulative distribution). The entries designated A are those that
follow directly from the photometric analysis. The entries designated
B are those that are drawn from Gillon et al.~(2008) and are repeated
here for convenience. The entries designated C are based on a
synthesis of our modeling results and theoretical models of stellar
evolution, as discussed in \S~4. The entries designated D are the
parameters of a refined transit ephemeris based on the two
newly-measured transit times as well as some other available timing
data (see \S~5).

As a consistency check we tried fitting the 2008~Aug~19 data and the
2008~Oct~09 data separately. We found that the results for the
parameters $R_p/R_\star$, $R_\star/a$, $i$, $u_1$, and $u_2$ were all
in agreement within 1$\sigma$, suggesting that our error estimates are
reasonable.

\begin{deluxetable}{lccc}
\tabletypesize{\scriptsize}
\tablecaption{System Parameters of WASP-4b\label{tbl:params}}
\tablewidth{0pt}

\tablehead{
\colhead{Parameter} & \colhead{Value} & \colhead{68.3\% Conf.~Limits} & \colhead{Comment}
}

\startdata
{\it Transit ephemeris:} & & & \\
Reference epoch~[BJD]                                 & $2454697.797562$  &  $\pm 0.000043$    & D  \\
Orbital period~[days]                                 & $1.33823214$      &  $\pm 0.00000071$  & D  \\
& & & \\
{\it Transit parameters:} & & & \\
Midtransit time on 2008~Aug~19~[BJD]                  & $2454697.797489$  &  $\pm 0.000055$    & A  \\
Midtransit time on 2008~Oct~09~[BJD]                  & $2454748.650490$  &  $\pm 0.000072$    & A  \\
Planet-to-star radius ratio, $R_p/R_\star$             & $0.15375$         &  $-0.00055$, $+0.00077$ & A  \\
Orbital inclination, $i$~[deg]                        & $88.56$           &  $-0.46$, $+0.98$  & A  \\
Scaled semimajor axis, $a/R_\star$                     & $5.473$           &  $-0.051$, $+0.015$  & A  \\
Transit impact parameter, $b = a\cos i/R_\star$        & $0.143$           &  $-0.098$, $+0.038$ & A  \\
Transit duration~[hr]                                 & $2.1660$          &  $-0.0062$, $+0.0054$ & A  \\
Transit ingress or egress duration~[hr]               & $0.2913$          &  $-0.0002$, $+0.01006$ & A  \\
Linear limb-darkening coefficient, $u_1$                & $0.311$           &  $\pm 0.041$  & A \\
Quadratic limb-darkening coefficient, $u_2$             & $0.227$           &  $\pm 0.089$  & A \\
Semimajor axis~[AU]                                   & $0.02340$         &  $\pm 0.00060$ & C \\
Planet-to-star mass ratio, $M_p/M_\star$               & $0.00127$        &  $\pm 0.00012$  & C \\
& & & \\
{\it Stellar parameters:} & & & \\
Mass, $M_\star$~[M$_{\odot}$]                           &  $0.925$         &  $\pm 0.040$  & C  \\
Radius, $R_\star$~[R$_{\odot}$]                         &  $0.912$         &  $\pm 0.013$  & C  \\
Surface gravity, $\log g_\star$~[cgs]                 &  $4.4813$        &  $\pm 0.0080$ & C  \\
Mean density, $\rho_\star$~[g~cm$^{-3}$]               &  $1.728$         &  $-0.047$, $+0.016$   & A  \\
Effective temperature, $T_{\rm eff}$~[K]               &  $5500$          &  $\pm 100$      & B  \\
Metallicity, [Fe/H]                                  &  $-0.03$         &  $\pm 0.09$    & B  \\
Projected rotation rate, $v\sin i_\star$~[km~s$^{-1}$]  &  $2.0$          &  $\pm 1.0$     & B  \\
Luminosity [L$_\odot$]                                &  $0.682$          &  $\pm 0.065$ & C  \\
Absolute $V$ magnitude                                &  $5.30$          &  $\pm 0.13$  & C \\
Age [Gyr]                                             &  $6.5$          &  $\pm 2.3$  & C  \\
& & & \\
{\it Planetary parameters:} & & & \\
$M_p$~[M$_{\rm Jup}$]                                 &  $1.237$          &  $\pm 0.064$ & C \\
$R_p$~[R$_{\rm Jup}$]                                 &  $1.365$         &  $\pm 0.021$ & C \\
Surface gravity, $g_p$~[m~s$^{-2}$]                   &  $16.41$         &  $\pm 0.75$ & A  \\
Mean density, $\rho_p$~[g~cm$^{-3}$]                  &  $0.604$          &  $\pm 0.042$ & C \\
\enddata

\tablecomments{ (A) Based on the analysis of the new light curves (see
  \S~3). (B) From Gillon et al.~(2008). (C) Functions of group A and B
  parameters, supplemented by theoretical Y$^2$ isochrones (see
  \S~4). For (C) parameters, the error bars are based only on the
  measurement errors and do not account for any possible error in the
  Y$^2$ isochrones. (D) Based on the two newly-measured transit times,
  as well as all entries in Table~2 of Gillon et al.~(2008).}

\end{deluxetable}

\section{Theoretical isochrone fitting}

The combination of transit photometry and the spectroscopic orbit
(radial-velocity variation) of the host star do not uniquely determine
the masses and radii of the planet and the star. There remain fitting
degeneracies $M_p \propto M_\star^{2/3}$ and $R_p \propto R_\star
\propto M_\star^{1/3}$ (see, e.g., Winn 2008). We broke these
degeneracies by requiring consistency between the observed properties
of the star, the stellar mean density $\rho_\star$ that can be derived
from the photometric parameter $a/R_\star$ (Seager \& Mallen-Ornelas
2003, Sozzetti et al.~2007), and theoretical models of stellar
evolution. The inputs were $T_{\rm eff}=5500\pm 100$~K and
[Fe/H]~$=-0.03\pm 0.09$ from Gillon et al.~(2008), the stellar mean
density $\rho_\star = 1.694_{-0.037}^{+0.017}$~g~cm$^{-3}$ derived
from the results for the $a/R_\star$ parameter, and the Yonsei-Yale
(Y$^2$) stellar evolution models by Yi et al.~(2001) and Demarque et
al.~(2004). We computed isochrones for the allowed range of
metallicities, and for stellar ages ranging from 0.1 to 14 Gyr. For
each stellar property (mass, radius, and age), we took a weighted
average of the points on each isochrone, in which the weights were
proportional to $\exp(-\chi^2_\star/2)$ with
\begin{equation}
\chi^2_\star =
\left[ \frac{\Delta{\rm [Fe/H]}}{\sigma_{{\rm [Fe/H]}}} \right]^2 +
\left[ \frac{\Delta T_{\rm eff}}{\sigma_{T_{\rm eff}}} \right]^2 +
\left[ \frac{\Delta \rho_\star}{\sigma_{\rho_\star}} \right]^2.
\end{equation}
Here, the $\Delta$ quantities denote the deviations between the
observed and calculated values at each point. The asymmetric error bar
in $\rho_\star$ was taken into account by using different values of
$\sigma_{\rho_\star}$ depending on the sign of the deviation. The
weights were further multiplied by a factor taking into account the
number density of stars along each isochrone, assuming a Salpeter mass
function. This procedure is essentially the same as that employed by
Torres et al.~(2008). The only difference is that we calculated the
68.3\% uncertainties by assuming that the errors in $T_{\rm eff}$,
$\rho_\star$, and [Fe/H] obey a Gaussian distribution, while Torres et
al.~(2008) took the distribution to be uniform within the quoted
1$\sigma$ limits.

Through this analysis, we found $M_\star = 0.925 \pm 0.040$~$M_\odot$
and $R_\star = 0.912 \pm 0.013$~$R_\odot$. The stellar age was poorly
constrained, with a formally allowed range of $6.5\pm 2.3$~Gyr and a
nearly uniform distribution. The corresponding planetary mass and
radius were obtained by merging the results for the stellar properties
with the parameters determined in our photometric analysis, and with
the stellar radial-velocity semiamplitude $K_\star = 0.24\pm
0.01$~km~s$^{-1}$ measured by Wilson et al.~(2008). The results are
$M_p = 1.237\pm 0.064$~$M_{\rm Jup}$ and $R_p = 1.365 \pm
0.021$~$R_{\rm Jup}$. Table~\ref{tbl:params} gives these results,
along with the values for some other interesting parameters that can
be derived from the preceding results. As a consistency check, we
computed the implied stellar surface gravity and its uncertainty based
on our analysis, finding $\log g = 4.481 \pm 0.008$ where $g$ is in
cm~s$^{-2}$. This agrees with the spectroscopic determination of
surface gravity, $\log g = 4.3 \pm 0.2$, made by Wilson et al.~(2008)
based on an analysis of the widths of pressure-sensitive lines in the
optical spectrum.

It is important to keep in mind that the quoted error bars for the
parameters designated C in Table~\ref{tbl:params} are based on the
measurement errors only, and assume that the any systematic errors in
the Y$^2$ isochrones are negligible. As a limited test for the
presence of such errors, Torres et al.~(2008) tried analyzing transit
data for 9 systems using isochrones computed by three different
groups: the Y$^2$ isochrones used here as well as those of Girardi et
al.~(2000) and Baraffe et al.~(1998). They found the differences to be
smaller than the error bars, especially for stars similar to the Sun
such as WASP-4. Similar results were found by
Southworth~(2009). Nevertheless, the true systematic errors in the
Y$^2$ isochrones are not known, and we have not attempted to quantify
them here, although it seems plausible that the masses and radii are
subject to an additional error of a few percent.

\section{Transit Times and a Refined Ephemeris}

Precise measurements of transit times are important because the
gravitational perturbations from other bodies in the system (such as a
satellite or additional planet) could produce detectable variations in
the orbital period. Based on the MCMC analysis described in the
previous section, the uncertainties in the two transit times are 4.6~s
and 4.9~s, making them among the most precise such measurements that
have been achieved. This is a consequence of good photometric
precision and fine time sampling, along with the relative
insignificance of time-correlated noise and the large transit depth.

As a check on the error bars, we also estimated the midtransit time
and its error using the residual permutation (RP) method, a type of
bootstrap analysis that attempts to account for time-correlated
errors. Fake data sets are created by subtracting the best-fitting
model from the data, then adding the residuals back to the model after
performing a cyclic permutation of their time indices. For each fake
data set, $\chi_F^2$ is minimized as a function of $T_c$, $\Delta
m_0$, $k_z$ and $k_t$ (the other parameters being held constant, as
they are nearly uncorrelated with $T_c$). The distribution of the
results for $T_c$ is taken to be the {\it a posteriori}\, probability
distribution for $T_c$. For the 2008~Aug~19 data, the RP-based error
bar was only 4\% larger than the MCMC-based error bar; the results
were nearly indistinguishable. For the 2008~Oct~09 data, the RP-based
error was 25\% larger than the MCMC-based error. To be conservative,
we report the larger RP-based errors in Table~1 and used those larger
error bars in recomputing the ephemeris (see below).

As a further check on the analysis, we allowed TS1 and TS2 to have
independent values of $T_c$, $\Delta m_0$, $k_z$, and $k_t$ during the
fitting process. In other words we fitted all the data but did not
require that TS1 and TS2 agree on the transit time. The result was
that the difference between $T_c$(TS1) and $T_c$(TS2) was $2.3 \pm
13.4$~s. We also tried a similar experiment with the 2008~Aug~19 data,
splitting it into two equal parts that were fitted jointly with TS1
and TS2 (which in this case were required to agree on the midtransit
time so as to provide a constraint on the transit duration). The
result in this case was $\Delta T_c = 4.6\pm 11.9$~s. The mutual
consistency of the results suggests that our error bars are
reasonable.

The transits of 2008~Aug~19 and 2008~Oct~09 were separated by
38~orbital periods. By calculating $\Delta T_c / 38$ we derive an
independent estimate of the orbital period, $P=1.3382369\pm
0.0000024$~days. The most precise determination previously reported
was $P=1.3382324\pm 0.0000029$~days, by Gillon et al.~(2008), based on
2 years of data. Our period has comparable precision, though it is
based on only 2 transits separated by 50~days. The difference between
the two independent period determinations is $0.39\pm 0.32$~s.

We fitted a linear function of epoch to the two newly-measured
midtransit times along with the 5 transit times given in Table~2 of
Gillon et al.~(2008). The fit had $\chi^2$=7.8 with 5 degrees of
freedom. The chance of finding a value of $\chi^2$ this large by
chance is 17\%, using the quoted 1$\sigma$ error bars and assuming the
errors obey a Gaussian distribution. We deem this an acceptable fit,
and conclude that the available data do not provide compelling
evidence for any departures from a constant period.  The refined
transit ephemeris is $T_c(E) = T_c(0) + EP$, with
\begin{eqnarray}
T_c(0) & = & 2,454,697.797562 \pm 0.000043 {\rm [BJD],} \nonumber \\
P      & = & 1.33823214 \pm 0.00000071.
\label{eq:ephemeris}
\end{eqnarray}
Fig.~2 shows a plot of the differences between the observed and
calculated transit times.

\begin{figure}[p]
\epsscale{1.0}
\plotone{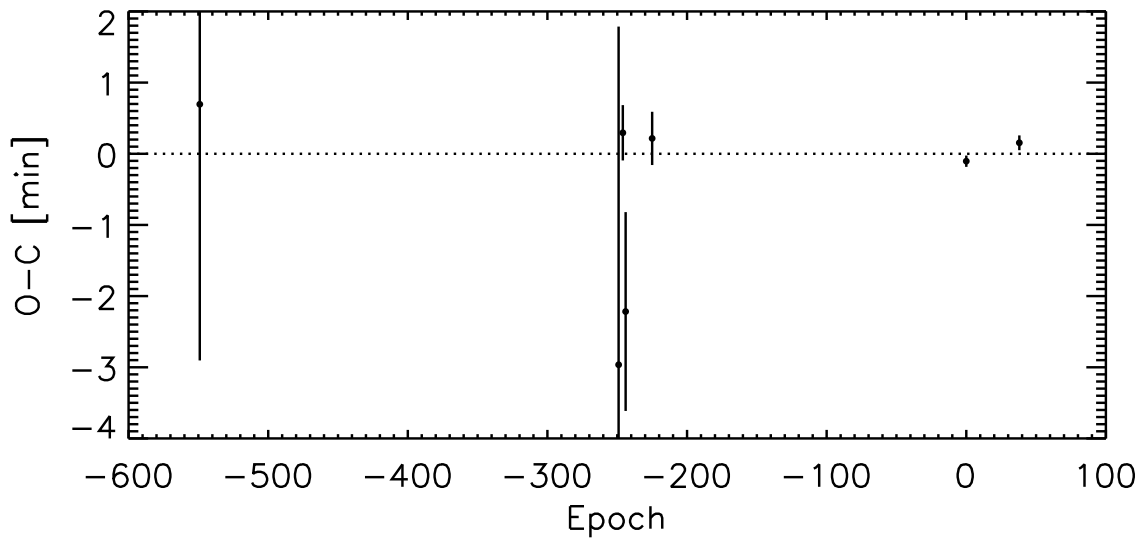}
\caption{ Transit timing residuals for WASP-4b. The calculated times
  (using the ephemeris given in Eq.~\ref{eq:ephemeris}) have been
  subtracted from the observed times.
\label{fig:o-minus-c}}
\end{figure}

\section{Discussion}

Our results for the planetary and stellar properties are in accord
with the previous analyses by Wilson et al.~(2008) and Gillon et
al.~(2008). In general our error bars are comparable in size or
smaller than those of Gillon et al.~(2008), who observed a transit
with one of the Very Large Telescopes. This consistency is reassuring,
especially since our error estimates are more conservative in some
respects. We have accounted for uncertainty in the limb-darkening
coefficients, as well as the slopes of systematic trends with time and
airmass. Previous investigators assumed that these parameters were
known exactly, leading to underestimated errors in any covariant
parameters. Southworth~(2008) demonstrated this effect for the
limb-darkening coefficients in particular.\footnote{Our results for
  the limb-darkening coefficients are $u_1 =0.311\pm 0.041$ and
  $u_2=0.227\pm 0.089$. These are not too far from the tabulated
  values $0.265$ and $0.303$ given by Claret~(2004) for a star with
  the observed effective temperature and surface gravity. However,
  with data this precise, one should not fix the limb-darkening
  coefficients at the tabulated values. The tables have unquantified
  errors and these errors are strongly correlated with the other
  photometric parameters. For example, for HD~209458 the tabulated
  values can be ruled out with $>$99.9\% precision (Southworth~2008).}
In addition, Gillon et al.~(2008) found a much stronger influence of
correlated noise on the results for the midtransit time. In the
language of \S~3, their RP-based error bar was 300\%--400\% larger
than the MCMC-based error bar, as compared to 4--25\% in our case. We
also question the applicability of the ``small-planet'' approximation
used by Gillon et al.~(2008). With our data, using the small-planet
approximation would have led to an erroneous 2.3$\sigma$ shift in
$R_p/R_\star$.

For the key parameter $R_p$, Wilson et al.~(2008) found
$1.416_{-0.043}^{+0.068}$~$R_{\rm Jup}$, Gillon et al.~(2008) found
$1.304_{-0.042}^{+0.054}$~$R_{\rm Jup}$, and we find $1.365\pm
0.021$~$R_{\rm Jup}$. Here it is important to reiterate that the
estimate of $R_p$ relied on the Y$^2$ isochrones, which probably
contribute an additional systematic error of a few percent. Despite
this, it seems clear that WASP-4b is a member of the class of
``bloated'' planets, by which we mean planets that are larger than
predicted according to theoretical models of solar-composition gas
giant planets, even after accounting for the intense irradiation from
the parent star.

Bodenheimer et al.~(2003) predicted the radii of giant planets as a
function of the age, mass, and equilibrium temperature, defined as
\begin{equation}
T_{\rm eq} = \left[ \frac{(1-A) L_\star}{16\pi\sigma a^2} \right]^{1/4} =~~(1677~{\rm K})~(1-A)^{1/4},
\end{equation}
where $A$ is the Bond albedo, $L_\star$ is the stellar luminosity,
$\sigma$ is the Stefan-Boltzmann constant, and $a$ is the semimajor
axis. In the latter equality we have evaluated $T_{\rm eq}$ for
WASP-4b using the results given in Table~\ref{tbl:params}. As long as
the albedo is not very close to unity, Bodenheimer et al.~(2003)
predict a planetary radius of 1.13~$R_{\rm Jup}$ for a solar
composition at 4.5~Gyr. This is smaller than the observed radius by
11$\sigma$.

Fortney et al.~(2007) presented theoretical models parameterized by
mass, age, and an effective orbital distance, defined as the distance
from the Sun where a hypothetical planet would receive the same flux
as the actual planet,
\begin{equation}
a_\oplus = a \left( \frac{L_\star}{L_\odot} \right)^{-1/2} =~~0.0281~{\rm AU},
\end{equation}
where again we have evaluated the expression as appropriate for
WASP-4b. For a solar composition at 4.5~Gyr, Fortney et al.~(2007)
predict $R_p = 1.16$~$R_{\rm Jup}$, about 10$\sigma$ smaller than the
observed value. Neither of these sets of models takes into account the
``transit radius effect'' of Burrows et al.~(2003)---the
underprediction of transit radii by modelers who use an
inappropriately high pressure to define the radius of their
models---but this effect is expected to be only a few percent
according to Fortney et al.~(2007).

One might suppose that the system is young, and has therefore not had
as much time to contract from its presumably hot and distended initial
state. For an age of 300~Myr, Fortney et al.~(2007) predict a radius
of $1.30$~$R_{\rm Jup}$, only 3$\sigma$ smaller than the observed
value (and perhaps consistent, given the systematic errors in the
isochrones and the transit radius effect). However, such a young age
is disfavored by our isochrone analysis, and Wilson et al.~(2008) have
also argued that the star is older than about 2~Gyr based on a
nondetection of Li~I~($6708$~\AA) in the spectrum.

Another possible resolution is to invoke an extra source of internal
heat (of unspecified origin) within the planet. Liu et al.~(2008) have
provided useful fitting formulas to gauge how much extra power is
required, for a solar-composition planet of a given mass and effective
orbital distance. They are parameterized by the variable $x$, defined
as the ratio of the extra power to the incident power from the host
star's radiation. By interpolating Table~3 of Liu et al.~(2008) we
find that to match the observed radius of WASP-4b, one needs an
``anomalous'' heating rate of $\sim$$0.15$\% of the incident power
from the host star, or $\sim$$8\times 10^{26}$~erg~s$^{-1}$.
According to the same tables, the equilibrium radius is achieved in
$\sim$30~Myr.

Of course the preceding calculation does not solve the problem of the
bloated planet. It merely quantifies the power requirement for the
unknown heat source. The calculations of Liu et al.~(2008) were in the
context of tidal heating due to orbital circularization. According to
their Eq.~(A3), an orbital eccentricity of $0.002~(Q_p/10^5)$ would
suffice to produce enough power to inflate the planet to the requisite
degree. Hence for $Q_p$ between $10^5$ and $10^6$, the required
eccentricity for this hypothesis is about $0.002$ to $0.02$.  This is
small enough to be compatible with the RV data, but larger than
expected based on the estimated stellar age and the order-of-magnitude
timescale for tidal circularization (see \S~3), unless there is a
third body whose gravitational influence is preventing
circularization. A very precise measurement of the time of occultation
of WASP-4b by its host star might allow the orbital eccentricity (or
more precisely $e\cos\omega$, where $\omega$ is the argument of
pericenter) to be measured well enough to test this notion. Of course,
many other ways to explain the radii of the bloated planets have been
given in the literature (Guillot \& Showman 2002, Showman \& Guillot
2002, Bodenheimer et al.~2003, Burrows et al.~2007, Chabrier \&
Baraffe 2007). This issue remains unresolved for the particular case
of WASP-4b and for the entire ensemble.

In addition to refining the estimates of planetary parameters, and
seeking additional bodies through transit timing, an important
application of precise differential photometry is checking for
transits of planets that have been detected by the Doppler technique
but for which the orbital inclination is unknown. Given the progress
of the Doppler surveys in finding lower-mass planets, it is also a
likely route to the discovery of the first examples of ``super-Earth''
planets that transit bright stars, and that are therefore valuable for
understanding the sizes, orbits, and atmospheres of terrestrial
planets. It is generally thought that detection of transits of
terrestrial-mass planets must be done from space, to eliminate the
contaminating effects of Earth's atmosphere and to observe for long
intervals without interruptions due to bad weather and the day/night
cycle. These are clear advantages but they must be weighed against the
high cost of space missions.

It is interesting to contemplate how small a planet we would have been
able to detect, given data of the quality presented here. Our light
curve has a signal-to-noise ratio of approximately $\sqrt{\Gamma
  T}\delta/\sigma \approx 920$, using the notation of Carter et
al.~(2008) in which $\Gamma T$ is the number of data points per
transit duration, $\delta$ is the transit depth, and $\sigma$ the
photometric precision. Statistically one might expect that we would
have achieved a 5$\sigma$ detection of a planet that has an area
$920/5=184$ times smaller than WASP-4b, i.e., a planet with
$R_p=1.1$~$R_\oplus$. However, it is not clear whether a statistically
unassailable 5$\sigma$ detection would really be credible, given the
possible presence of systematic trends, correlated errors, uncertainty
in predicted transit times, and general caution.

To judge the believability of a super-Earth detection, we added to the
residuals of Fig.~2 a model light curve of a transiting super-Earth.
All of the parameters of the model were the same as the actual WASP-4b
parameters except the planetary radius, which was reduced to
1.75~$R_\oplus$, the approximate upper limit that is predicted for
terrestrial planets. The result, shown in Fig.~3, gives a visual
impression of what one might realistically expect, based on two nights
of observations of a 12th magnitude star with a large telescope.

We fitted this simulated data in nearly the same way that we fitted
the WASP-4 data. The only difference is that we fixed the values of
the limb-darkening parameters and the stellar radius, as befits a
detection rather than a characterization experiment. We also assumed
the orbital period was known to within 10\% (3.2~hr). The free
parameters were $(R_p/R_\star)$; $i$; the two mid-transit times; and
the parameters describing systematic trends, $\Delta m_0$, $k_z$ and
$k_t$, for each time series. The results were that $(R_p/R_\star)$ was
consistent with the input value and had an error of 14\%, dominated by
the correlation with the poorly-constrained orbital inclination. The
transit times were recovered within 5~min. The detection is visually
apparent in the time-binned light curve shown in the lower panel of
Fig.~3. One would certainly prefer a longer stretch of out-of-transit
data, especially if the orbital period has a large
uncertainty. Nevertheless the impression given is that the detection
of transits of super-Earths around Sun-like stars is within reach of
ground-based observations.

\begin{figure}[p]
\epsscale{1.0}
\plotone{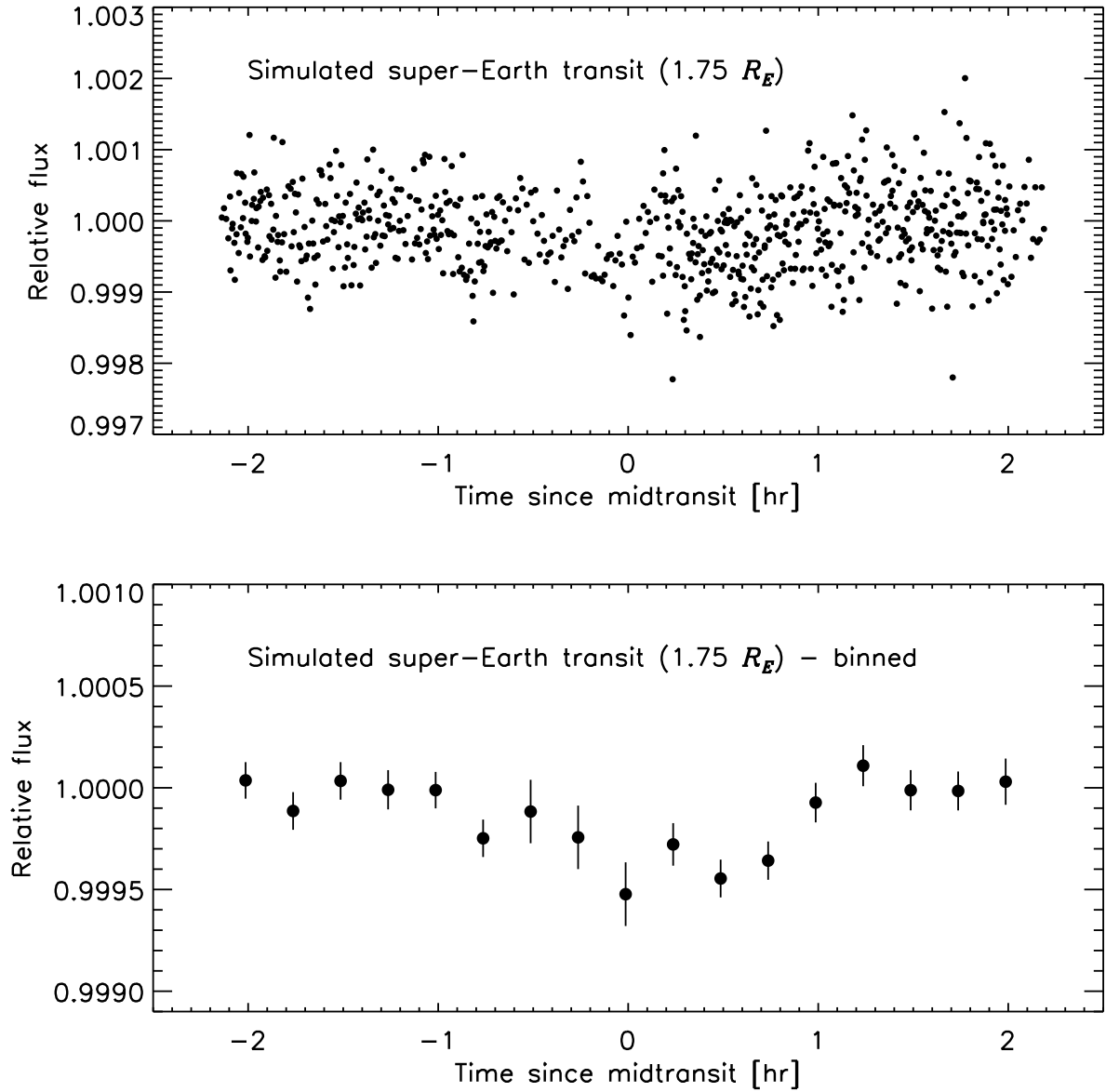}
\caption{Simulated photometric observations of a transiting
super-Earth around a 12th magnitude G7V star,
based on our WASP-4 data. The residuals shown in
Fig.~1 were added to an idealized transit model in
which $R_p=1.75~R_\oplus$ and all other parameters
were the same as in the WASP-4 model. The bottom panel
shows a time-binned version of the simulated data.
\label{fig:fakedata}}
\end{figure}

\acknowledgments We thank Paul Schechter and Adam Burgasser for
observing on our behalf on several occasions, for this project and
related projects, and for helpful discussions on observing
techniques. Partial support for this work came from NASA Origins
grants NNX09AB33G (to M.J.H.) and NNG04LG89G (to G.T.).

\end{document}